\documentclass[superscriptaddress,floatfix,nofootinbib,preprintnumbers]{revtex4}

\usepackage{amsmath,graphicx,times,footmisc}

\newcommand{\gsm}{\Gamma^\text{SM}}

\newcommand{\gi}{\Gamma^\text{inv}}
\newcommand{\gv}{\Gamma^\text{vis}}

\newcommand{\ifb}{\text{fb}^{-1}}

\begin{document}

\title{LHC: Standard Higgs and Hidden Higgs}

\author{Christoph Englert} 
\affiliation{Institute for Particle Physics
  Phenomenology, Department of Physics, Durham University, United
  Kingdom}

\author{Tilman Plehn} 
\affiliation{Institut f\"ur Theoretische Physik,
  Universit\"at Heidelberg, Germany}

\author{Michael Rauch} 
\affiliation{Institut f\"ur Theoretische Physik,
  Karlsruhe Institute of Technology, Germany}

\author{Dirk Zerwas}
\affiliation{LAL, IN2P3/CNRS, Orsay, France}

\author{Peter M. Zerwas} 
\affiliation{Deutsches Elektronen-Synchrotron
            DESY, Hamburg, Germany} 
\affiliation{Institut f\"ur Theoretische
            Teilchenphysik und Kosmologie, 
            RWTH Aachen University, Germany}

\date{\today}

\preprint{IPPP/11/81} 
\preprint{DCPT/11/162}
\preprint{KA-TP-39-2011} 
\preprint{SFB/CPP-11-77} 

\begin{abstract}
  \noindent {\it Interpretations of Higgs searches critically involve
    production cross sections and decay probabilities for different
    analysis channels. Mixing effects can reduce production rates,
    while invisible decays can reduce decay probabilities.  Both
    effects may transparently be quantified in Higgs systems where a
    visible Higgs boson is mixed with a hidden sector Higgs boson.
    Recent experimental exclusion bounds can be re-interpreted in this
    context as a sign for non-standard Higgs properties. Should a
    light Higgs boson be discovered, then our analysis will quantify
    how closely it may coincide with the Standard Model.}
\end{abstract}

\maketitle

\section{Basics}

\noindent
Recent experimental searches for Higgs bosons~\cite{Higgs} by
ATLAS~\cite{HiggsATLAS} and CMS~\cite{HiggsCMS}, based on luminosities
of up to about 2.3~$\ifb$, have excluded the Higgs boson of the
Standard Model, when combined~\cite{Rol}, between 141~GeV and 476~GeV
at 95\% CL.  ATLAS and CMS have updated several channels using an
integrated luminosity of up to 5~$\ifb$~\cite{ATLASCouncil,
  CMSCouncil}, so that each experiment alone has about the same
sensitivity as the combination in the low mass region, raising the
low-mass bound of the Higgs from the LEP2 limit \cite{LEP2} of 
114.4~GeV to 115.5/115~GeV and leading to an
exclusion of the mass from 131~GeV/127~GeV to 453~GeV/600~GeV by
ATLAS/CMS, respectively.  Bounds as low as fractions $\mathcal{R} \leq
0.3$ have been set on the production cross sections with respect to
the Standard Model [SM] for some of the masses probed. Given the far
reaching consequences, the question arises naturally to what extent
Higgs bosons, in slightly generalized scenarios, may still exist in
this mass region.

On the other hand, a gap from about 130~GeV down to the low-mass limit 
of about 115~GeV is left, presently, in which a SM-type Higgs signal can be
expected to either rise up or plunge in the near
future~\cite{ATLASCouncil, CMSCouncil}.  When a Higgs boson will
indeed be discovered in this mass range, the standard-hidden Higgs
scenario allows to quantify how well the global properties of the
particle coincide with the predictions of the Standard Model.

Both these questions have been addressed~\cite{Bock,Englert} at the
theoretical level in a standard-hidden Higgs scenario. The present
sequel will include the recent LHC results on the Higgs sector in the
analysis. The ground for the fundamental idea to use the Higgs boson
as a portal to a novel hidden sector in nature has been laid for
general model structures in Ref.~\cite{Higgs.portal}.  Phenomenological
implications~\cite{pheno} can then be studied in specific frameworks
like hidden valleys~\cite{hiddenvalley}.  Other than the Higgs portal
there exist two additional portal-type interactions in a
renormalizable theory, namely kinetic $U(1)$ mixing~\cite{mixing} and
mixing with sterile neutrinos~\cite{neutrinos}.

The experimentally observed rates depend on the Higgs production cross
sections and the decay branching ratios~\cite{book}. In scenarios in
which the standard Higgs boson is mixed with another Higgs boson,
originating from a hidden sector for instance, the production cross
sections are reduced universally by the mixing parameter and, in
addition, the visible branching ratios may be lowered by decays into
invisible channels induced by mixing. Throughout this paper we will
consider the minimal Standard Model Higgs sector as the activator of
the visible Higgs boson~\cite{rge}, but the same arguments trivially
hold for extended visible Higgs sectors, for example in supersymmetry
or other extensions to the Standard Model~\cite{review}.  In the early
theoretical analyses of Refs.~\cite{Bock,Englert} the two effects have
been joined properly, expanded here by exploiting data in addition.

Specific examples have also been discussed in Ref.~\cite{Wells}.
The opening of invisible channels has been investigated in
several other studies, motivated by a large variety of models ranging 
from Higgs decays to LSP's in supersymmetry \cite{Janot} and other SM 
extensions, see {\it e.g.} \cite{Shrock}, to phenomenological approaches 
\cite{invis.higgs,invis.exi}, partly connected with cold dark matter 
analyses~\cite{invis.models}. However, the effect of mixing on the 
production cross sections has often been left aside.

The Higgs potential of the standard-hidden Higgs system consists of
the standard potential in the visible sector and the isomorphic
potential in the hidden sector, coupled by a bi-bilinear portal
term~\cite{Higgs.portal}. Vacuum expectation values of the Higgs
fields build up the Higgs mass parameters in the two sectors as well
as their coupling. Diagonalizing the standard-hidden Higgs mass matrix
[squared]{\footnote{The essential elements of the formalism
    \cite{Bock,Englert} are summarized briefly in this introductory
    section as to render the letter self-contained for the reader's
    convenience.}}, 
generates the mass eigenvalues $M_1^2 < M_2^2$ and
rotates the standard-hidden Higgs states $H_s,H_h$ to the mass
eigen-states $H_1,H_2$,
\begin{alignat}{5}
  H_1 &=&  \cos\chi \, H_s + \sin\chi \, H_h &    \notag \\
  H_2 &=& -\sin\chi \, H_s + \cos\chi \, H_h & \,,
\label{eq:mixi}
\end{alignat}
the labels ordered parallel to the rising mass values. In the limit of
small mixing, $H_1$ approaches the lighter of the states $H_s$ or
$H_h$, {\it i.e.} $\cos\chi$ or $\sin\chi \to 1$, respectively, and
$H_2$ the heavier of the two primary states.

The mixing reduces the production cross sections of $H_1,H_2$ to
\begin{equation}
  \sigma_{1,2}  = \cos^2\chi \, \{\sin^2\chi\} \, \sigma^\text{SM}_{1,2}
\end{equation}
with respect to the cross sections of the Standard Model for
equivalent masses, the expressions within the curly brackets for $index = 2$
substituting the corresponding expressions for $index = 1$. 
The visible and invisible partial widths are modified correspondingly,
\begin{eqnarray}
  \gv_{1,2}  &=& \cos^2\chi \, \{\sin^2\chi\} \, \gsm_{1,2}
                 + \Delta^{vis}_2 \, \Gamma^\text{HH}_2     \nonumber \\
  \gi_{1,2}  &=& \sin^2\chi \, \{\cos^2\chi\} \, \Gamma^\text{hid}_{1,2}
                 + \Delta^{inv}_2  \, \Gamma^\text{HH}_2
\end{eqnarray}
are built up by decays to particles in the hidden sector, including 
cascade decays with $\Delta^{vis\{inv\}}_2 = \zeta^2_1 \, \{[1-\zeta_1]^2\}
\neq 0$ only for $index = 2$. After factorizing off the mixing parameters, the 
relevant $H_h$ widths $\Gamma^\text{hid}_{1,2}$, defined for virtual masses 
$M_{1,2}$, are generated by the unknown dynamics of the hidden sector. Most
agnostically, these width parameters should not exceed the values of
the Higgs masses, $\Gamma^\text{hid}_{1,2} \leq M_{1,2}$.  The partial
width $\Gamma^\text{HH}_2$ accounts for potential $H_2 \to H_1 H_1$
cascade decays if $M_2 > 2 M_1$~\cite{Wells} with the $H_1$ visible
decay branching ratio $\zeta_1 = 1/[1+\tan^2\chi \,
\Gamma_1^\text{hid}/ \Gamma^\text{SM}_{\text{tot},1}]$. In this case
also a mixed contribution $\Gamma^\text{HH,mix}_{2} = 2 \, \zeta_1 \,
[1-\zeta_1] \, \Gamma^\text{HH}_2$ of one $H_1$ decaying visibly and
the other one invisibly is present; numerical details have been
presented in Ref.~\cite{Englert}. Likewise the total widths,
\begin{equation}
  \Gamma^\text{tot}_{1} = \cos^2\chi \, \{\sin^2\chi\} \, \gsm_{\text{tot};1,2} 
  + \sin^2\chi \, \{\cos^2\chi\} \, \Gamma^\text{hid}_{1,2}
  + \Delta_2 \, \Gamma^\text{HH}_2               \,,                
\end{equation}
with $\Delta_2 = 0,1$ for $index = 1,2$, respectively, 
are free parameters at the phenomenological level in this scenario.

\section{Higgs measurements {\it vs} Higgs parameters}

\noindent
The search for Higgs bosons at the LHC by ATLAS~\cite{HiggsATLAS} and
CMS~\cite{HiggsCMS} has excluded major parts of the intermediate Higgs
mass range and beyond~\cite{Rol,ATLASCouncil,CMSCouncil}.  However, 
it leaves a clear gap down to the low-mass limit set originally by LEP2
and improved by ATLAS/CMS to about 115~GeV in which a Higgs signal 
may be confirmed or refuted in the coming
year~\cite{ATLASCouncil,CMSCouncil}.  In this Section we follow two
scenarios. First, we study Higgs sectors with reduced couplings and
invisible decay widths, which can naturally explain a missing Higgs
signal at the LHC. In a second step, we describe the effects of a
potential Higgs discovery in the still open mass window with SM-like
couplings.

\subsection*{Bounds}

\noindent To begin with, in the Higgs mass range ruled out for
Standard Model couplings, the corresponding Higgs cross sections are
constrained to values as low as $\mathcal{R} \sim 0.3$ in units of the
SM value. These bounds, as well as production cross sections measured
after the discovery of the Higgs boson, can be re-interpreted within
the standard-hidden Higgs scenario. For a given mass value the bounds
induce constraints on the mixing and on the invisible width of $H_1$,
while at the same time the second, heavier state of the Higgs pair is
proven not to violate the experimental bounds. The bounds on the
production cross sections for $H_1$ from any set of final states $F$
induce the following constraints at a given mass value on the mixing
parameter and the invisible width:
\begin{equation}
  \dfrac{\sigma[pp \to H_1 \to F]}{\sigma[pp \to H_1 \to F]^\text{SM}} 
  = \dfrac{\cos^2\chi}{1 + \tan^2\chi \, [{\Gamma^\text{hid}_1}/{\Gamma^\text{SM}_{\text{tot},1}}]}
  \leq \mathcal{R}                                                                            \,.
\label{eq:excl}
\end{equation}
[This observable coincides with the twin ratio $\Gamma_\text{p}
\Gamma_\text{d} / \Gamma_\text{tot}$ of Refs.~\cite{Bock,Englert},
composed of partial widths for the production and decay channels as
well as the total decay width.] Likewise, the non-observation of the
Higgs boson $H_2$ for the same mass value considered, with, this time,
the lighter companion of the Higgs pair hiding underneath the
experimental bounds for the final-state production rates, leads to a
similar formula with $\cos^2\chi$ and $\sin^2\chi$ interchanged and
potential cascade corrections supplemented.

When a Higgs boson will be discovered in the present gap down to the
LHC improved LEP2 limit, the bound on $\mathcal{R}$ will be replaced by the
measured value $\mathcal{R}$ of the ratio of observed to SM cross
section, to be explored later.
 
\begin{figure}[b]
  \begin{center}
    \includegraphics[height=0.3\textwidth]{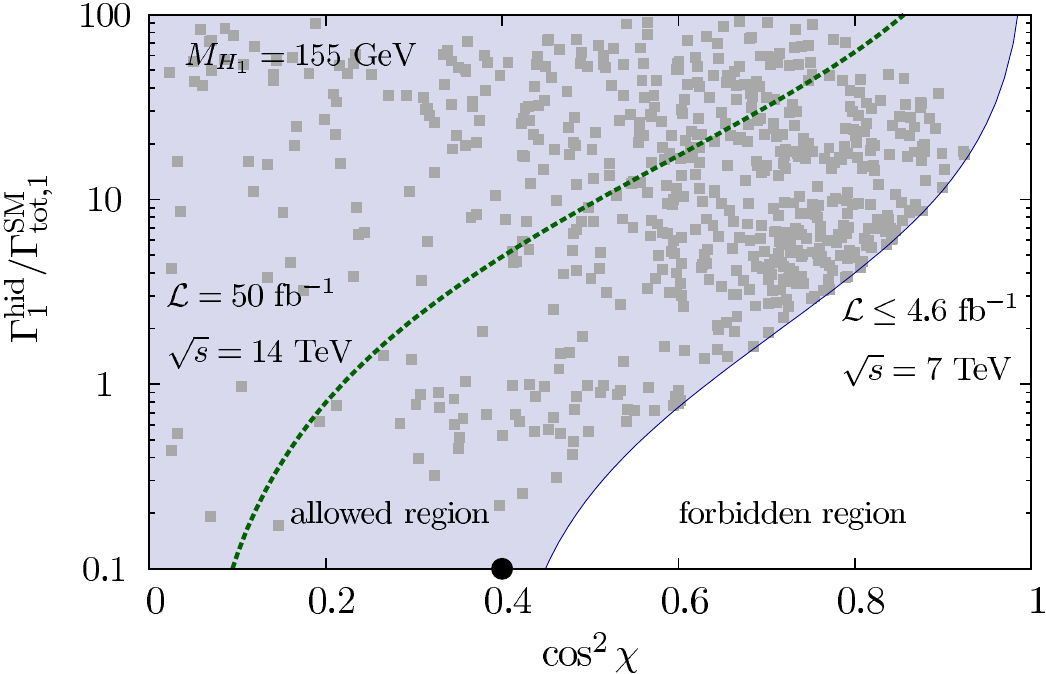}
  \end{center}
  \vspace*{-5mm}
  \caption{ \label{fig:cont_limit} \it{Bounds on the mixing and hidden
      decay width of $H_1$ for the point $M_{H_1} = 155$ $\text{GeV};
      \mathcal{R} = 0.4$ in the standard-hidden Higgs scenario, based
      on current experimental
      results~\cite{HiggsATLAS,HiggsCMS,Rol}. The regions dappled by
      small squares are compatible with unitarity and precision
      measurements. The dot indicates the $\Gamma_1^\text{hid} \to 0$
      limit of the exclusion curve at $\mathcal{R}$. The dotted line
      indicates the projected search limit for $\mathcal{L} =
      50~\ifb$. The corresponding results for $H_2$ with the same mass
      follow from mirroring all lines at $\cos^2\chi = 1/2$.}}
\end{figure}

Higgs decays into the hidden sector are invisible. This mode however
plays quite an important role for hidden-sector scenarios in
general. Even though the experimental analysis will be very demanding
at hadron colliders~\cite{invis.higgs,invis.exi}, less at lepton
colliders, the central points should nevertheless be analyzed
theoretically in the present context. In a first step, the cross
section for $H_1$ production with invisible decays may be bounded by
\begin{equation}
  \dfrac{\sigma[pp \to H_1 \to inv]}{\sigma[pp \to H_1]^\text{SM}}
  = \dfrac{\sin^2\chi \, [\Gamma^\text{hid}_1 / \Gamma^\text{SM}_{\text{tot},1}]}
  {1 + \tan^2\chi \, [{\Gamma^\text{hid}_1}/{\Gamma^\text{SM}_{\text{tot},1}}]}
  \leq \mathcal{J}                                                            \,.
\label{eq:inv}
\end{equation}
Analogously, the relation for $H_2$ follows immediately by exchanging
$\cos^2\chi$ with $\sin^2\chi$ and adding the cascade corrections
again.  Thus, the total of the visible channels plus the invisible
channel describes, globally, the essential elements of the
standard-hidden Higgs scenario.\bigskip

Derived from the inequality Eq.~{\ref{eq:excl}} above, current
experimental exclusion bounds on $\mathcal{R}$ give rise to lower
limits on the invisible decay widths in association with the mixing
parameters:
\begin{equation}
  \dfrac{\Gamma^\text{hid}_1}{\Gamma^\text{SM}_{\text{tot},1}} \geq \cot^2\chi \,  
  \left[\dfrac{\cos^2\chi}{\mathcal{R}} - 1 \right]                         \,
\label{eq:GamH}
\end{equation}
for $\cos^2\chi \geq \mathcal{R}$, while the bound is zero below;
similarly for $H_2$ by interchanging $\cos^2\chi$ and $\sin^2\chi$ and
adding the corrections possibly for cascade decays.

To evaluate the experimental Higgs bounds, we have chosen the
characteristic point $\{M_H = 155\, \text{GeV}; \mathcal{R} = 0.4 \}$
within the broad shoulder between 155~GeV and 175~GeV, where the
observed $\cal R$ limit is essentially
constant~\cite{HiggsATLAS,HiggsCMS,Rol}.  Identifying $H$ first with
$H_1$ and following Eqs.~\ref{eq:excl},~\ref{eq:GamH}, the area above
the contour in Fig.~\ref{fig:cont_limit} is still viable for the
potential observation of a Higgs boson with increased luminosity at
the LHC.  As expected, the exclusion contours move up with reduced
$\mathcal{R}$.  The corresponding areas for $H$ identified with $H_2$
probed for the same mass are just mirrored at $\cos^2\chi = 1/2$.  For
$\mathcal{R} < 1/2$, the $H_1$ and $H_2$ curves cross each other at a
positive value of
${\Gamma^\text{hid}}/{\Gamma^\text{SM}_{\text{tot}}}$, leaving a blank
area between the exclusion contours not covered by either Higgs boson.

The ratios $\Gamma^\text{hid} / \Gamma^\text{SM}_\text{tot}$ may run
from small up to high values since the Higgs-boson widths in the
standard and the hidden sectors may be quite different. The areas
dappled by little squares, generated by random points in the space of
vacuum expectation values and couplings, is compatible with
constraints from unitarity and high-precision measurements, {\it cf.}
Ref.~\cite{Englert}. Small values of $\cos^2\chi$ effectively amount
to heavy Higgs scenarios and are disfavored by electroweak precision
measurements. The squares nevertheless cover almost the whole allowed
parameter space.  Note, however, that, when a light $H_2$ is probed,
the density of points compatible with the bounds for light $H_1$
production \cite{lephwg} is reduced.

The small dot denotes the point where the exclusion bound ends after
touching the $\cos^2\chi$-axis at $\mathcal{R}$. For values on the
half-axis $\cos^2\chi$ larger than $\mathcal{R}$ a non-vanishing value
of $\Gamma^\text{hid}$ is necessary to still be compatible with the
experimental bound.  In the area below the dotted line the Higgs
production cross section~\cite{Hprod,Hprod2,Hprod3,Hprod4,Hprod5} in
association with different numbers of jets~\cite{jet_veto} is large
enough to be accessed experimentally for a luminosity of $50~\ifb$.

If an upper bound $\mathcal{J}$ on the invisible cross section can be
established, Eq.~{\ref{eq:inv}}, an upper bound on the Higgs width in
the hidden sector follows for $\cos^2\chi \geq \mathcal{J}$:
\begin{equation}
  \dfrac{\Gamma^\text{hid}_1}{\Gamma^\text{SM}_{\text{tot},1}} \leq
  \dfrac{\cot^2\chi \; \mathcal{J}}{\cos^2\chi - \mathcal{J}}        \,,
\end{equation}
with, correspondingly, $\cos^2\chi \Leftrightarrow \sin^2\chi$ for $1
\Leftrightarrow 2$ plus cascade contributions. The contour is
illustrated in Fig.~{\ref{fig:cont_disc}}, with the 68\% CL region at
50 $\ifb$ luminosity and a center-of-mass energy of 14 TeV given by
the shaded area. The shaded area reflects the intrinsic theoretical
uncertainties on the perturbative Higgs production cross section [such
as pdf and scale uncertainties] as well as the expected experimental 
uncertainty at the fixed integrated luminosity. All these effects are included in
SFitter~\cite{sfitter}, which we employ to compute the uncertainty
band in Fig.~{\ref{fig:cont_disc}} [and later analogously in
Fig.~\ref{fig:contSM}].

In total, $H_1$ or $H_2$ Higgs bosons with mixing and hidden-width
parameters in the allowed regions can, presently, still be realized in
the standard-hidden Higgs scenario.

\begin{figure}[htb]
  \begin{center}
    \includegraphics[height=0.3\textwidth]{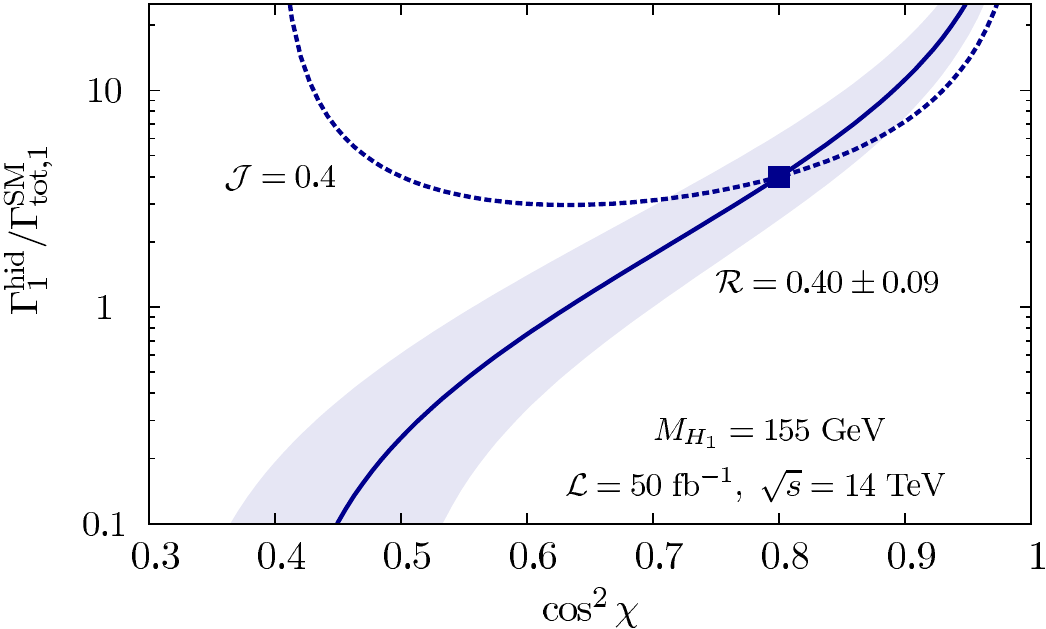}
  \end{center}
  \vspace*{-5mm}
  \caption{ \label{fig:cont_disc} \it{Contours for mixing and hidden
      decay width of the Higgs boson $H_1$ if visible and invisible
      ratios of cross sections are measured.  The uncertainties of 
      $\mathcal{R}$, shown at 68\% CL as shaded band, 
      are determined by means of SFitter~\cite{sfitter} including both 
      experimental and theoretical errors. 
      The uncertainties of $\mathcal{J}$ cannot be reliably estimated
      at the present time. A currently not excluded
      Higgs mass of 155~GeV with $\mathcal{R} = \mathcal{J} = 0.4$ is
      chosen for illustration.  The crossing point defines the central
      values of both parameters. The analysis for $H_2$ runs parallel
      after replacing $\cos^2\chi \rightarrow \sin^2\chi$.}}
\end{figure}

\subsection*{Measurements}

\noindent 
Should a light Higgs boson be discovered at the LHC, the measurement
of $\mathcal{R}$ provides a first global picture of the physical
nature of the state. Instead of just constraining parameter areas, the
mixing angle and the width in the hidden sector can then be correlated
by exploiting Eq.~{\ref{eq:GamH}}, reduced now to equality. In the
limit $\mathcal{R} \to 1$ the interpretation of the observed state is
unique as the $H_1$ contour in Fig.~{\ref{fig:contSM}}, for $H_1$
identified with the state, shrinks to the SM corner $\{\cos^2\chi \to
1, \Gamma^\text{inv}_1=\sin^2\chi\,\Gamma^\text{hid}_1 \to 0$\}.  When
identifying the observed state with $H_2$, the $H_2$ contour leads to
analogous conclusions in the equivalent limit $\sin^2\chi \to 1$.
Note that for small mixing the two states $H_1$ and $H_2$, identified
consecutively with the state experimentally observed, approach the
same Higgs state $H_s$ of the Standard Model, as pointed out
earlier. Thus the SM Higgs boson emerges in a model-independent way
from the experimental data. The agreement with the Standard Model can
quantitatively be described by the size of the contour curves
approaching the SM corners.

The twin sister of the observed state is difficult to detect
experimentally in the SM corner. If the observed state is identified
with $H_1$, for instance, the production of the higher-mass $H_2$
sister is strongly suppressed in the SM approach for an $H_1$ mass
chosen as $M_{H_1} = 125$~GeV and several $H_2$ masses;
exemplified for a typical set of parameters:
\begin{center}
\begin{tabular}{cc|cc|cc|cc|cc}
  \hline
  ${\mathcal{R}}_1$  \; & ${\mathcal{R}}_2 $ & $M_{H_1}$ [GeV]& $M_{H_2}$ [GeV]& 
  $\cos^2\chi$ & $\sin^2\chi $ &  $\Gamma_1^\text{hid} / \Gamma_1^\text{SM}$ &
  $\Gamma_2^\text{hid} / \Gamma_2^\text{SM}$ 
  & $\sigma_{H_1}$ [pb] & $\sigma_{H_2}$ [pb] \\
  \hline
  \hline
  0.7 & 0.05  & 125  & 155 &  0.85 & 0.15  & 1.21  & 0.35 & 48.5  &5.9 \\
  &           &         & 300 &      &    & & & & 1.9\\
  &           &         & 450 &      &    &  & & & 1.3\\
  \hline     
\end{tabular} 
\end{center}
Even for a Higgs boson mass of 400~GeV, the best combined
limit~\cite{Rol} at $\mathcal{R} \sim 0.3$ is not yet sensitive.
Thus, the properties of $H_1$ must be exploited primarily for studying
the nature of the standard-hidden Higgs sector.

If the invisible parameter $\mathcal{J}$ can be measured
experimentally, the crossing of the $\mathcal{J}$ with the
$\mathcal{R}$ contour,~{\it cf.} Fig.~\ref{fig:cont_disc}, fixes the two physical parameters at the point
marked by the square,
\begin{equation}
  \cos^2\chi = \mathcal{R}_1 + \mathcal{J}_1 \quad \text{and} \quad
  \dfrac{\Gamma^\text{hid}_1}{\Gamma^\text{SM}_\text{tot,1}} = 
  \dfrac{\mathcal{J}_1}{\mathcal{R}_1}\,\dfrac{\mathcal{R}_1+\mathcal{J}_1}
  {1-(\mathcal{R}_1+\mathcal{J}_1)}
\end{equation}
for the observed state identified with the Higgs state $1$, and
correspondingly $\sin^2\chi$ instead of $\cos^2\chi$ if the observed
state is identified with state $2$.  The result would invite the
search for the second Higgs boson of the standard-hidden pair in
either case. Depending on the value of $\cos^2\chi$, the wave function
of the Higgs boson $H_1$, given by $\cos\chi$ and $\sin\chi$, will
determine the primary affiliation either with the standard Higgs boson
$H_s$ or the hidden Higgs boson $H_h$, and complementary for $H_2$.
Note, however, that the nearly parallel running of the $\mathcal{R}$
and $\mathcal{J}$ curves for large $\cos^2\chi$ in Fig.~\ref{fig:cont_disc} 
renders the measurement of
the crossing point experimentally very difficult when the errors on
$\mathcal{J}$ are included.

We conclude by observing that the two Higgs masses $M_{1,2}$ and the
ratios $\mathcal{R}_{1,2}$ and $\mathcal{J}_{1,2}$ measured for
$H_{1,2}$ complete, in general, the parameter determination of this
theory, {\it i.e.} the elements of the Higgs matrix and the hidden
widths of the Higgs field $H_s$ at the respective physical Higgs mass
values. In fact, the measurement of all the ratios $\mathcal{R}_{1,2}$
and $\mathcal{J}_{1,2}$ over-determines the parameter set, and the sum
rule $\mathcal{R}_1 + \mathcal{J}_1 + \mathcal{R}_2 + \mathcal{J}_2 =
1$ can be exploited to test the consistency of the standard-hidden
Higgs scenario with the experimental results.

\begin{figure}[htb]
  \begin{center}
    \includegraphics[height=0.3\textwidth]{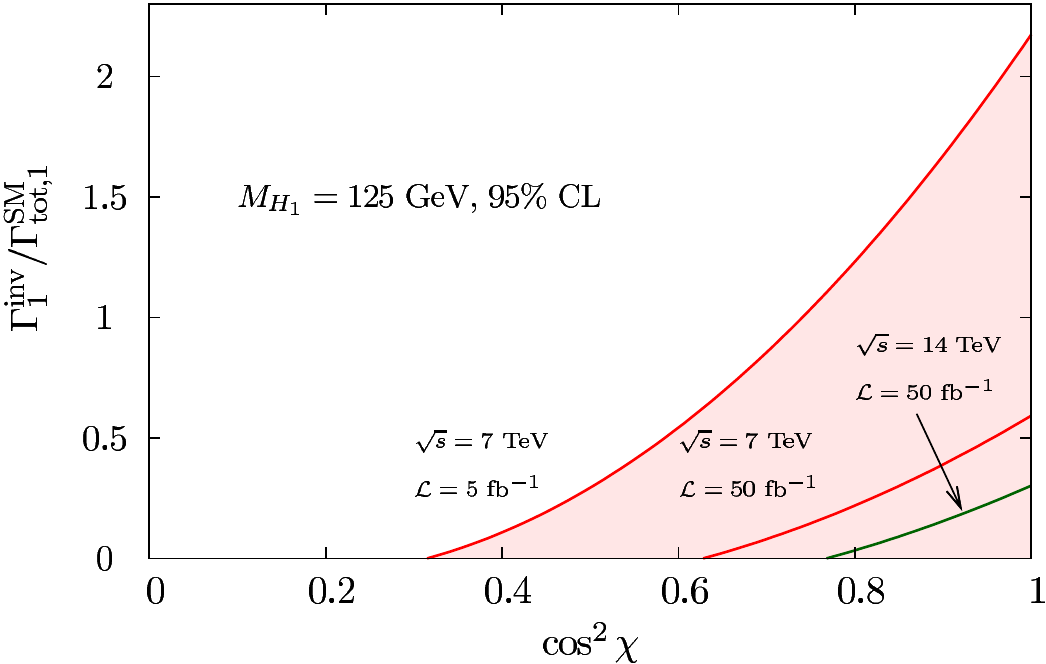}
  \end{center}
  \vspace{-5mm}
  \caption{ \label{fig:contSM} \it{95\%~CL contours for the mixing and
      hidden decay width from the observation of a SM-like Higgs boson
      [$\mathcal{R}=1$] of mass 125~GeV.  The uncertainty bands for
      different LHC energies and luminosities are determined by means 
      of SFitter~\cite{sfitter} including both 
      experimental and theoretical errors.
      To focus on the SM limit proper, $\{\cos^2\chi,\Gamma^\text{inv}_1\}
      \to \{1,0\}$, the observable $\Gamma^\text{inv}_1$ is chosen as parameter 
      in the figure.}}
\end{figure}

\section{Conclusions}

\noindent
Expanding on earlier theoretical analyses, we have exploited recent
LHC data to draw conclusions on the Higgs sector in a standard-hidden
Higgs scenario which develops continuously out of the Standard
Model. Coupling the Standard Model with a hidden sector suggests an
exciting interpretation of the recent experimental data.\bigskip

-- Mixing effects and invisible decays of Higgs particles can reduce
the cross sections for experimental searches of the Higgs boson at the
LHC. Though the SM Higgs boson may be ruled out for specific mass
values, if cross sections fall below the SM predictions, Higgs bosons
are not excluded {\it per se} for these mass values.  Constraints then
follow for the mixing of standard and hidden Higgs states, as well as
for decays into the hidden sector. Such Higgs bosons should emerge
however with rising LHC luminosity. \bigskip

-- If a Higgs boson is discovered at the LHC, the standard-hidden
Higgs scenario allows to quantify how closely the state coincides with
or to what extent it deviates from the Standard Model. It is
demonstrated in Fig.~\ref{fig:contSM} how from present collider
parameters such a measurement will rapidly gain sensitivity with
increasing luminosity and energy. \bigskip

\noindent
Thus, the standard-hidden Higgs system not only describes an exciting
physical scenario, but it can also be adopted for interpreting the
data in searches for Higgs bosons at the LHC and, if discovered, for
the global understanding of their nature.

\acknowledgments

\noindent
C.~Englert and P.~M.~Zerwas are grateful to Heidelberg University for
the warm hospitality extended to them during several visits.
C.~Englert acknowledges funding by the Durham International Junior
Research Fellowship scheme. M.~Rauch acknowledges support by the
Deutsche Forschungsgemeinschaft via the
Sonderforschungsbereich/Transregio SFB/TR-9 ``Computational Particle
Physics'' and the Initiative and Networking Fund of the Helmholtz
Association, contract HA-101 (``Physics at the Terascale'').


\end{document}